\documentclass[twocolumn,showpacs,preprintnumbers,amsmath,amssymb]{revtex4}

\usepackage{mathrsfs}
\usepackage{txfonts}

\usepackage{graphicx}
\usepackage{dcolumn}
\usepackage{bm}
\usepackage{slashed}

 {\catcode`\|=\active \gdef|{\egroup\,\vrule\,\bgroup}}

\def\be{\begin{eqnarray}}
\def\ee{\end{eqnarray}}
\begin{document}


\title{Electron Scattering from Freely Moveable spin-$\frac{1}{2}$ fermion in Strong Laser Field}

\author{Ai-Hua Liu  and Shu-Min Li}%
\affiliation{Department of Modern Physics,  University of Science
and Technology of China,  P.  O.  Box 4,  Hefei,  Anhui 230026,
  People's Republic of China}


\date{\today}

\begin{abstract}
{We study the electron scatter from the freely movable
spin-$\frac{1}{2}$ particle in the presence of a linearly polarized
laser field in the first Born approximation. The dressed state of
electrons is described by a time-dependent wave function derived
from a perturbation treatment (of the laser field). With the aid of
numerical results we explore the dependencies of the differential
cross section on the laser field properties such as the strength,
the frequency, as well as on the electron-impact energy, etc. Due to
the targets are movable, the DCS of this process reduced comparing
to the Mott scattering, especially in small scattering angles.}
\end{abstract} \vskip 0.5in

PACS number(s): 34.80.Qb, 34.80.Dp, 32.80.Wr. \pacs {number(s):
34.80.Qb, 34.80.Dp, 32.80.Wr.}
 \vskip 1in
\maketitle
%

%
%
Physics related to the radiative processes experienced by free
electrons inside a strong electromagnetic field were studied since
the advent of laser sources in the early 1960's
\cite{Cite01,Cite02}. An overview on this field can be found in the
textbooks by Mittleman \cite{Mitt93} and Fedorov \cite{Fedo97} and
some other recent reviews \cite{Fran90,Ehlo98,Ehlo01}. Most of these
studies are carried out in the regime of non-relativistic collisions
and for low- or moderate-field intensities. There are also some
studies have been carried out to investigate theoretically the
relativistic potential. In the presence of ultrastrong lasers, a
relativistic treatment becomes imperative (even for slow electrons).
In the treatments of Refs.\cite{Ehlo88,Kami99,Pane99} , effects
related to the electron spin have been neglected and the electron
has been considered as a Klein-Gordon particle. Based on the theory
of Refs.\cite{Rosh,Deni}, Szymanowski $et~al.$ \cite{Szym97,Szym98}
investigated the spin effect in the relativistic potential
scattering in the presence of a circularly polarized field, however,
as they stated, the resulting expression for circularly polarized
field turned out to be more tractable than for the general case of
elliptical or linear polarizations, and then
 Li $et~al.$ for the case of linearly polarized field \cite{Lism03} and
 Attaourti $et~al.$ for the cases of circularly and elliptically polarized fields \cite{Atta}.
 Manaut $et~al.$ investigated the case of polarized electrons \cite{Mana05}.

The present study addresses the problem of an electron scattering
off the freely moveable target in the presence of a monochromatic
linearly polarized homogeneous laser field. The aim of this study is
to add some physical insight and to show the modification of
differential cross section (DCS) due to the movable target and
compare to the case of Mott scattering. We investigate, to be
specific, the relativistic scattering of an electron from freely
movable proton/positive-muon and its recoil effect. A differential
cross section is derived by the trace procedure with the aid of
Feynman Calcu, Mathematica, and a simplified form is given for
specific.
%
Unless specifically stated, atomic units (a.u.) $\hbar=m=e=1$ are
used throughout, and the matric tensor is
$g^{\mu\nu}=diag(1,-1,-1,-1)$.

%

%
    In the regime of laser field intensity as considered in this paper,
the field can be treated as classically that does not allow pair
creation \cite{Bula96}, its four-potential that satisfies the Lorenz
condition $\partial A(x)=0$ is described by(linear polarization):
   $ {A(x)=a\cos(kx)} \label{eq:as},$
where a=(0,$\textbf{a}$), and $\textbf{a}$ is the amplitude of
vector potential of the field. The four wave vector of field is $k$
=(${\omega}$/c, ${\textbf{k}}$), where $\omega$ and $\emph{k}$ being
the frequency and wave number, respectively.

This scattering process  involves two fermions, including an
electron($e^{-}$), a proton ($p$) or positive-muon ($\mu^{+}$).
\begin{equation}
{p(\kern 0.3ex \mu^{+})+e^{-} \rightarrow p( \kern 0.3ex
\mu^{+})+e^{-}.} \label{eq:process}
\end{equation}

The relativistic, asymptotic electron state in laser field can be
described by Dirac-Volkov function \cite{Volkov35}, when normalized
in volume V, considering the linearly polarized field, it reads:
\begin{equation}
{\psi_q}(x)={\psi_p}(x)=\left(1+\frac{\slashed{k}\slashed{A}}{2c(kp)}\right)\frac{u(p,s)}{\sqrt{2QV}}e^{iS(x)}.
\label{eq:q}\end{equation}
\begin{equation}
S(x)=-qx-\frac{a\cdot p}{c(k\cdot p)}\sin(kx), \label{eq:sx}
\end{equation}
where $u$ represents a bispinor for the free electron which us
normalized as $\overline{u}u=2c^{2}$, and ${{q}}^{\mu}=(Q/c,
{\textbf{q}})$ is the averaged four-momentum of electron in the
presence of the laser field
%
${q}^{\mu}=p^{\mu}-\frac{\overline{a^{2}}}{2c^{2}(k\cdot
p)}k^{\mu}.$
%
where $\overline{a^{2}}$ is time-averaged square of the
four-potential of the laser field. The square of this momentum
$q^{\mu}$ is ${q}^{\mu}q_{\mu}=m^{2}_{*}c^{2}$. The parameter
$m_{*}=\sqrt{1-\frac{a^{2}}{c^{4}}}$ is an effective mass of the
electron in the radiative field

In a first approximation proton/posiive-muon will be treated as a
structureless, spin-1/2 Dirac particle. Comparing to the electron,
it is much heavier, therefore its modification in the presence of
laser is not so notable and we can use plane wave $\psi_{P}(x)$ to
describe it.

Then the S-matrix element for scattering process takes the form
\begin{widetext}
\begin{equation}
 {S_{fi}=-i\int
 d^{4}x \overline{\psi}_{q_{f}}(x) \slashed{A}(x) \psi_{q_{i}}(x)}=i\int
 d^{4}x d^{4}y \overline{\psi}_{q_{f}}(x)\gamma^{\mu}\psi_{q_{i}}(x)D_{F}(x-y) \overline{\psi}_{P_{f}}(y)
 \gamma_{\mu}\psi_{P_{i}}(y).
 \label{eq:sm1}
\end{equation}
\end{widetext}
with $D_{F}$ is the Feynman propagator for electromagnetic
radiation.

For a useful result we calculated the $d\sigma$ in the laboratory
frame of reference in which the initial proton/muon is at rest and
set $q_{f}=(Q^{'},\textbf{q}^{'}),q_{i}=(Q,\textbf{q}_{i})$ and
$P_{i}=(Mc,\textbf{0})$. To get the unpolarized cross section we
must average over initial states and sum over final ones. Then using
the relation $d^{3}q_{f}=\frac{1}{c^{2}}q^{'}Q^{'}dQ^{'}d\Omega^{'}$
and integrating over the final state energy $Q_{f}$, we get for the
scattering DCS \cite{Bjor64}:
\begin{widetext}
\begin{equation}
\frac{d\overline{\sigma}}{d\Omega^{'}}=\sum_{l}\frac{d\overline{\sigma}^{l}}{d\Omega^{'}}=\frac{1}{(2\pi)^{2}}\frac{1}{16Mc^{6}}
\sum_{l} \frac{{\bf q^{'}}}{{\bf
q}_{i}}\frac{1}{q^{4}}\frac{\overline{|M}^{(l)}_{fi}|^{2}}{Mc^{2}+Q+l\omega-
\frac{Q^{'}}{c^{2}q^{'}}(q_{i}\cos\theta+lk\sin\theta\cos\phi)},
\label{eq:dsdw}
\end{equation}
where
\begin{equation}
\label{eq:m2average}
|\overline{M}^{(l)}_{fi}|=\frac{1}{4}\sum_{l=-\infty}^{\infty}Tr[(c\slashed{p}_{f}+c^{2})
\Gamma^{\mu}_{(l)}(c\slashed{p}_{i}+c^{2})\overline{\Gamma}^{\nu}_{(l)}]
Tr[(c\slashed{P}_{f}+c^{2})\gamma_{\mu}^{(l)}(c\slashed{P}_{i}+c^{2})\gamma_{\nu}^{(l)}],
\end{equation}
\end{widetext}
with
$\Gamma^{\mu}=\triangle_{0}\gamma^{\mu}+\triangle_{1}\gamma^{\mu}\slashed{k}\slashed{a}+\triangle_{2}\gamma^{\mu}\slashed{a}\slashed{k}+
\triangle_{3}\gamma^{\mu}\slashed{a}\slashed{k}\gamma^{\mu}\slashed{k}\slashed{a}.$
Here the $\Delta_0,\Delta_1,\Delta_2,$ and $\Delta_3,$ have  the
same meaning as those of Ref.~\cite{Lism03}.

The energy conservation derived from the $\delta$-function is
\begin{equation}
Q^{'}(Mc^{2}+Q+l\omega)-c^{2}(\textbf{q}^{'}\cdot
\textbf{q}_{i}+l\textbf{q}^{'}\cdot
\textbf{k})=m_{*}^{2}c^{4}+Mc^{2}Q+lc^{2}(M\omega
-\textbf{q}_{i}\cdot \textbf{k}). \label{eq:e_conserve}
\end{equation}

In the limit case of $Q\ll Mc^2$, the scattering potential is
approach the fixed Coulomb potential, and there is laser field
assisted. The energy conservation here become
\begin{equation}
Q^{'}=Q+l\omega \label{eq:qpw}
\end{equation}
This is exactly the Mott scattering that has been discussed in
Ref.~\cite{Lism03}.
%

%

In table~\ref{tab:table1} we display the laser-assisted $e-p$
scattering differential cross section (DCS) for laser-assisted at
the field strength ${\cal E}_{0}=1.0 \times 10^8 V/cm$ and photon
energy $\hbar \omega = 1.17$ eV, for both geometries  ${\cal
E}_{0}\perp \textbf{p}$ and ${\cal E}_{0}\parallel \textbf{p}$
respectively  (at the scattering angle $\theta= 90^{\circ}$ the
azimuthal angle is $\phi=0^{\circ}$). It is shown that the DCS for
scattering is greatly enhanced with the application of laser field.
Due to the target is moveable, there is
 small modification on the DCS, the cross section is a little
smaller that that of laser assisted Mott scattering\cite{Lism03}.
With the impact energy increasing, the modifications become larger.

\begin{table}
\caption{\label{tab:table1}The differential cross section (DCS) for
laser-assisted $e-p$ scattering at the field strength ${\cal
E}_{0}=1.0 \times 10^8 V/cm$, and frequency $\hbar \omega =1.17$ eV.
$Log_{10}\frac{d\sigma_{0}}{d\Omega}$: result for laser free;
$Log_{10}\frac{d\sigma_{Mott}}{d\Omega}$: result for Mott
scattering; $Log_{10}\frac{d\sigma_{\parallel}}{d\Omega}$: result
for ${\cal E}_{0}\parallel \textbf{p}$. }
\begin{ruledtabular}
\begin{tabular}{cccccc}
 $\theta(^{\circ})$ & $E_{T_i} $
& $Log_{10}\frac{d\sigma_{0}}{d\Omega}$ &
$Log_{10}\frac{d\sigma_{Mott}}{d\Omega}$ &
$Log_{10}\frac{d\sigma_{\parallel}}{d\Omega}$&
\\
\hline

 $90$ & $ 1$ &~-9.1035    & ~-7.0904   &~-7.0907   \\
 $90$ & $10$ &-10.9225    & ~-8.8490   &~-8.8515   \\
 $90$ & $20$ &-11.4943    & ~-9.4169   &~-9.4218   \\
 $90$ & $40$ &-12.0821    & -10.0000   & -10.0094   \\
 $90$ & $80$ &-12.6826    & -10.6378   & -10.6558   \\
 $90$ & $160$&-13.4947    & -11.2346   & -11.4219   \\
 $90$ & $320$&-13.9197    & -11.7881   & -11.8469   \\
 $180$ & $ 1$ &-10.1032    & ~-8.7878     &~-8.7879     \\
 $180$ & $10$ &-13.3037    & -11.9318     &-11.9282     \\
 $180$ & $20$ &-14.3907    & -13.0591     &-13.0139     \\
 $180$ & $40$ &-15.1959    & -14.2225     & -13.8190     \\
 $180$ & $80$ &-15.4298    & -15.4539     & -14.1010     \\
 $180$ & $160$&-15.5149    & -16.6473     & -14.1861     \\
 $180$ & $320$&-15.6363    & -17.7977     & -14.2591
\end{tabular}
\end{ruledtabular}
\end{table}



    In this work we study the electrons scattering from protons in the presence of a
  radiation field. The theoretical results for the linear polarization case show that
  the DCS of scattering is greatly enhanced by the presence of the laser field,
  and reduced compared the Mott scattering of the same situation. The treatment can be readily
  extended to the case of a general polarized of the field, and even the cases of M{\"u}ller scattering and Bhabha scattering.


    This work is supported by the National Natural Science Foundation of
China under Grant Numbers 10475070 and 10674125.

\clearpage
%


\begin{thebibliography}{s2}
%
    \bibitem{Cite01} L. S Brown and T. W. B. Kibble,Phys. Rev. {\bf
    133}, A705 (1965); T. W. B. Kibble, {\it ibid.} {\bf 150}, 1060
    (1966); 
    \bibitem{Cite02} I. I. Goldman,Phys. Lett. {\bf 8},103(1964);
    Zh.\'{E}ksp. Teor. Fiz. {\bf 46}. 1412(1964) [Sov. Phys. JETP {\bf 19}, 954(1964)]; A. I. Nikishov and V. I. Ritus, {\it ibid}. {\bf 46}, 1768(1964) [{\bf 19},119]; N. B. Narozhny, A. I. Nikishov and V. I. Ritus, {\it ibid}.{\bf 47},930(1964) [{\it ibid}. {\bf 20},622(1965)].
    \bibitem{Mitt93} M. H. Mittleman, {\it Introduction to the Theory of Laser-Atom Interactions} (Plenum, New York, 1993).
    \bibitem{Fedo97} M. V. Fedorov, {\it Atomic and Free Electrons in a Strong Light
    Field}(World Scientific, Singapore,1997).
    \bibitem{Fran90} P. Francken and C. J. Joachain, J. Opt. Soc. Am. B {\bf 7}, 554 (1990).
    \bibitem{Ehlo98} F. Ehlotzky, A. Jaro\'{n}, and J. Z. Kami\'{n}ski,
Phys. Rep. {\bf 297}, 63 (1998).
     \bibitem{Ehlo01} F. Ehlotzky, Phys. Rep. {\bf 345}, 175 (2001).
    %
    \bibitem{Ehlo88}F. Ehlotzky, Opt. Commun. {\bf 66}, 265 (1988).
    \bibitem{Kami99}J. Z. Kamin\'{s}ki and F. Ehlotzky, Phys. Rev. A {\bf 59}, 2105 (1999).
    \bibitem{Pane99}P. Panek, J.Z. Kamin\'{s}ki, and F. Ehlotzky, Can. J. Phys. {\bf 77}, 591
(1999)
    \bibitem{Deni} M.M. Denisov and M. V. Fedorov, Zh. {\'E}ksp. Teor. Fiz. {\bf 53},
1340(1967) [Sov. Phys. JETP {\bf 26}, 779(1968)].
    \bibitem{Rosh} S.P. Roshchupkin, Laser Phys. {\bf 3}, 414(1993); {\bf 7}, 873(1997);
Zh. Eksp. Teor. Fiz. {\bf 106}, 102(1994)[JETP {\bf 79}, 54(1994)];
{\bf 109}, 337(1996) [{\bf 82}, 177(1996)].
    \bibitem{Szym97} C. Szymanowski, V. V\'{e}niard, R. Ta{\"{\i}}eb, A. Maquet,
and C. H. Keitel, Phys. Rev. A {\bf 56}, 3846 (1997).
    \bibitem{Szym98} C. Szymanowski and A. Maquet, Opt. Express {\bf 2}, 262 (1998).
\bibitem{Lism03} S.-M. Li, J. Berakdar, J. Chen, and Z.-F. Zhou, Phys. Rev. A {\bf 67}, 063409 (2003).
    \bibitem{Atta} Y. Attaourti and B. Manaut, Phys. Rev. A {\bf 68}, 067401 (2003);Y. Attaourti, B. Manaut, and A. Makhoute, {\it ibid}. {\bf 69}, 063407(2004); Y. Attaourti and S. Taj, {\it ibid.} {\bf 69}, 063411 (2004); Y.Attaourti, B. Manaut, and S. Taj, {\it ibid.} {\bf 70}, 023404 (2004).
    \bibitem{Mana05} B. Manaut, S.Taj, Y. Attaourti, Phys. Rev. A {\bf 71}, 043401 (2005);

    \bibitem{Bula96} C. Bula $et$ $al.$, Phys. Rev. Lett. {\bf 76}, 3116 (1996).
    \bibitem{Volkov35} D. M. Volkov, Z. Phys. {\bf 94}, 250 (1935).
    \bibitem{Bjor64} J. D.Bjorken, S. D. Drell, ,{\it Relativistic Quantum
    Mechanics} (McGraw-Hill, New York, 1964)

    \end{thebibliography}
\end{document}